\documentclass[11pt]{article}

\usepackage{amsmath,amssymb}

\newcommand{\beq}{\begin{eqnarray}}
\newcommand{\eeq}{\end{eqnarray}}

\begin{document}

\begin{center}


\begin{flushright}
BCCUNY-HEP/06-05
\end{flushright}

\vspace{15pt}
{\large \bf CONFINEMENT IN 
$(2+1)$-DIMENSIONAL GAUGE THEORIES AT WEAK COUPLING }

\vspace{20pt}

{\bf Peter Orland}$^{\rm a.b.}$\!\footnote{giantswing@gursey.baruch.cuny.edu}

\vspace{8pt}

\begin{flushleft}
a. Physics Program, The Graduate School and University Center,
The City University of New York, 365 Fifth Avenue,
New York, NY 10016, U.S.A.
\end{flushleft}

\begin{flushleft}
b. Department of Natural Sciences, Baruch College, The 
City University of New York, 17 Lexington Avenue, New 
York, NY 10010, U.S.A. 
\end{flushleft}

\vspace{10pt}

{\bf Abstract}
\end{center}

\noindent 
In axial gauge, the $(2+1)$-dimensional SU($N$) Yang-Mills theory is
equivalent to a set of $(1+1)$-dimensional integrable models with a non-local 
coupling between charge densities. This fact makes it possible to determine
the static potential between charges at 
weak coupling in an anisotropic version of the theory and understand features of the spectrum. This 
note is based on a 
talk contributed
at {\em Quark Confinement and the Hadron Spectrum 7}, Sept. 2-7, 2006, Ponta Degada,
S\~ao Miguel, Azores.

\hfill

Many pictures and models have been proposed for confinement in QCD. Some 
sort of magnetic condensation clearly occurs, but no one knows why. Our 
grasp of the basic mechanism of confinement, at weak bare coupling, is little better than it was 
thirty years ago. In the opinion of the author, strong-coupling and variational methods can guide us 
toward
a better understanding, but are not 
substitutes for first-principles weak-coupling calculations. We discuss
such calculations here for a more modest theory. This is a $(2+1)$-dimensional SU($N$) gauge theory with
two coupling constants \cite{PhysRevD71}, \cite{PhysRevD74}, \cite{k-string}.

The action is of Yang-Mills type: $\int d^{3} {\mathcal L}$, where the Lagrangian is
${\mathcal L}=\frac{1}{2e^{\prime\,\,2}} {\rm Tr} F_{01}^{2} +
\frac{1}{2e^{2}} {\rm Tr} F_{02}^{2}
-\frac{1}{2e^{2}} {\rm Tr} F_{12}^{2}$,
where $A_{0}$,$ A_{1}$ and $A_{1}$ are SU($N$)-Lie-algebra-valued components of the gauge field,
and the
field strength is  $F_{\mu\nu}=\partial_{\mu}A_{\nu}-\partial_{\nu}A_{\mu}-i[A_{\mu},A_{\nu}]$. The
gauge transformation is 
$A_{\mu}(x) \rightarrow ig(x)^{-1}[\partial_{\mu}-iA_{\mu}(x)]g(x)$, where $g(x)$ is an SU($N$)-valued
scalar field. To study this model, we will take $e^{\prime}\ll e$; by doing this we lose rotation invariance. 

The next step is to discretize the $2$-direction, so that $x^{2}=a,2a,3a\dots$, where
$a$ is a lattice spacing. All fields will be considered functions of $x=(x^{0},x^{1}, x^{2})$. We define
the unit vector ${\hat 2}=(0,0,1)$. We 
replace $A_{2}(x)$ by a field $U(x)$ lying in
SU($N$), via $U(x)\approx \exp -iaA_{2}(x)$. There is a natural discrete covariant-derivative 
operator:
${\mathcal D}_{\mu}  {\mathfrak U}(x)
=\partial_{\mu}{\mathfrak U}(x)-{\rm i}A_{\mu}(x){\mathfrak U}(x)+{\rm i}{\mathfrak U}(x)A_{\mu}(x+{\hat 2}a)$,
$\mu=0,1$, for any $N\times N$ complex matrix field ${\mathfrak U}(x)$. The action is
$S=\int dx^{0} \int dx^{1}\sum_{x^{2}}\;a\;{\mathcal L}$ where
\beq
{\mathcal L}
=\frac{1}{2e^{\prime\,\,2}} {\rm Tr} F_{01}^{2} +
\frac{1}{2g_{0}^{2}}{\rm Tr} [{\mathcal D}_{0} U(x)]^{\dagger}
{\mathcal D}_{0}U(x) 
-\frac{1}{2g_{0}^{2}}{\rm Tr} [{\mathcal D}_{1} U(x)]^{\dagger}{\mathcal D}_{1}U(x) \;,
\label{sigmalagrangian}
\eeq
and $g_{0}^{2}=e_{0}^{2}a$. The Lagrangian 
(\ref{sigmalagrangian}) is invariant under the gauge transformation:
$A_{\mu}(x) \rightarrow ig(x)^{-1}[\partial_{\mu}-iA_{\mu}(x)]g(x)$ and
$U(x)\rightarrow g(x)^{-1}U(x)g(x+{\hat 2}a)$
where again, $g(x)\in {\rm SU}(N)$ and $\mu$ is restricted to $0$ or $1$. Notice 
that the quantity $g_{0}$ is dimensionless. In the limit $a\rightarrow 0$ , (\ref{sigmalagrangian}) yields
 the anisotropic continuum action.. The action (\ref{sigmalagrangian}) is 
a collection of parallel $(1+1)$-dimensional ${\rm SU}(N)\times {\rm SU}(N)$
sigma models, each of which couples to the gauge fields $A_{0}$, $A_{1}$. The sigma model field
is $U(x^{0},x^{1},x^{2})$, and each discrete $x^{2}$ corresponds to a different sigma model. The sigma-model self-interaction is the dimensionless number
$g_{0}$.

The left-handed and right-handed currents are, 
$j^{\rm L}_{\mu}(x)_{b}={\rm i}{\rm Tr}\,t_{b} \, \partial_{\mu}U(x)\, U(x)^{\dagger}$ and
$j^{\rm R}_{\mu}(x)_{b}={\rm i}{\rm Tr}\,t_{b} \, U(x)^{\dagger}\partial_{\mu}U(x)$, respectively, 
where $\mu=0,1$. The Hamiltonian obtained from (\ref{sigmalagrangian}) is $H_{0}+H_{1}$, where
\beq
H_{0}\!=\!\sum_{x^{2}}\int dx^{1} \frac{1}{2g_{0}^{2}}\{ [j^{\rm L}_{0}(x)_{b}]^{2}+[j^{\rm L}_{1}(x)_{b}]^{2}\}
\;,\label{HNLSM}
\eeq
and
\beq
H_{1}\!\!&\!\!=\!\!&\!\! \sum_{x^{2}}  \int dx^{1} \,
\frac{(g_{0}^{\prime})^{2}a^{2}}{4}\,\partial_{1}\Phi(x^{1},x^{2})\partial_{1}\Phi(x^{1},x^{2}) \nonumber \\
\!\!&\!\!-\!\!&\!\! 
\left(\frac{g_{0}^{\prime}}{g_{0}}\right)^{2}\,\,\sum_{x^{2}=0}^{L^{2}-a}  \int dx^{1} \!\!
\left[ j^{\rm L}_{0}(x^{1},x^{2})\Phi(x^{1},x^{2}) -j^{\rm R}_{0}(x^{1},x^{2}) \Phi(x^{1},x^{2}+a) \right]  
\nonumber \\
&+&(g_{0}^{\prime})^{2}q_{b}\Phi(u^{1},u^{2})_{b} -(g_{0}^{\prime})^{2}
q^{\prime}_{b}\Phi(v^{1},v^{2})_{b}   \; ,
\label{continuum-local}
\eeq
where $-\Phi_{b}=A_{0\,\,b}$ is the temporal gauge field, $g_{0}^{\prime\,\,2}=e^{\prime\,\, 2}a$, and
where in the last term
we have inserted two color charges - a quark with charge $q$ at site $u$
and an anti-quark with charge $q^{\prime}$ at site $v$. There is some gauge invariance left
over after the axial gauge fixing, namely that 
for each $x^{2}$
\beq
\left\{ \int d x^{1}\left[ j^{L}_{0}(x^{1},x^{2})_{b}-j^{R}_{0}(x^{1},x^{2}-a)_{b}\right] - g_{0}^{2}Q(x^{2})_{b} \right\}\Psi=0\;,
\label{physical}
\eeq
where $Q(x^{2})_{b}$ is the total color charge from quarks at $x^{2}$ and $\Psi$ is any physical 
state.

The Hamiltonian (\ref{HNLSM}), (\ref{continuum-local}) can be derived more carefully, by starting with 
the Kogut-Susskind lattice formulation \cite{PhysRevD71}, \cite{PhysRevD74} and assuming the lattice spacing
is small in the $x^{1}$-direction. In any case, we assume that $H_{1}$ is suitably regularized.

From (\ref{continuum-local}) we see that the left-handed charge of the sigma model
at $x^{2}$ is coupled to the electrostatic potential at $x^{2}$. The right-handed charge
of the sigma model is coupled to the electrostatic potential at $x^{2}+a$. The 
excitations of $H_{0}$, which we call Fadeev-Zamoldochikov or FZ particles, behave like solitons, though
they do not correspond to classical configurations. Some of these FZ particles 
are elementary and others are bound states of
the elementary FZ particles. An elementary FZ particle has an adjoint charge and mass $m_{1}$. An 
elementary FZ particle state
is a superposition of color-dipole states, with a quark  
charge at $x^{1}, x^{2}$ and an anti-quark charge at $x^{1},x^{2}+a$.  The interaction
$H_{1}$ produces a linear potential between color charges with the same value of $x^{2}$. Residual gauge
invariance (\ref{physical}) requires that at each value of $x^{2}$, the total color charge is zero. If there are 
no quarks, the total right-handed charge of FZ particles in the sigma model
at $x^{2}-a$ is equal to the total left-handed charge of FZ particles in the sigma model at $x^{2}$.

The presence of a mass gap and the lack of magnetization in the 
${\rm SU}(N)\times {\rm SU}(N)$ sigma model implies
confinement in the (2+1)-dimensional SU($N$) gauge theory in an anisotropic weak-coupling 
approximation $g_{0}\gg g_{0}^{\prime}$ \cite{PhysRevD71}. A formal 
weak-coupling perturbation theory in $g_{0}^{\prime}$ around the sigma-model states can
be considered in principle. Unfortunately, it is very hard to carry out this perturbation theory
in practice, except for gauge group SU($2$) \cite{PhysRevD74}.

The principal chiral sigma model spectrum is described by particles, each of which 
has a label $n$ which has the values $n=1,\dots,N-1$
\cite{pol-wieg}, \cite{abda-wieg}.  Each particle of label $n$ has an antiparticle 
of the same mass, with label $N-n$. The
masses are given by
\beq
m_{n}=m_{1}\frac{\sin\frac{n\pi}{N}}{\sin\frac{\pi}{N}},\;\; m_{1}=\frac{C}{a}(g_{0}^{2}N)^{1/2}e^{-\frac{4\pi}{g_{0}^{2}N}}+{\rm non\!-\!universal \;corrections}\;, \label{mass-spectrum}
\eeq
where $C$ is a non-universal constant. 
 
Lorentz invariance
in each $x^{0},x^{1}$ plane is manifest; hence the linear potential is not the only effect of
$H_{1}$. The interaction also creates and destroys pairs of elementary
FZ particles. This effect is unimportant, however, provided the interaction is small enough. This specifically means that the square of
the $1+1$ string tension in the $x^{1}$-direction coming from $H_{1}$ is small compared to
the square of the mass of fundamental FZ particle.

A rough picture of a gauge-invariant state for the gauge group SU($2$) with no quarks
is:
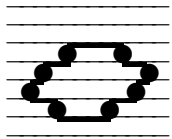
\begin{center}

\begin{picture}(150,60)(30,0)

\linethickness{0.5mm}

\multiput(75,-0.5)(0,7){8}{\multiput(0,0)(5,0){12}{\put(0,0){$-$}}}

\put(95,12){\circle*{7}}
\put(85,19){\circle*{7}}
\put(90,26){\circle*{7}}
\put(99,33){\circle*{7}}
\put(115,12){\circle*{7}}
\put(125,19){\circle*{7}}
\put(130,26){\circle*{7}}
\put(120,33){\circle*{7}}

\put(95,8.3){\line(1,0){20}}
\put(115,15.3){\line(1,0){10}}
\put(125,22.3){\line(1,0){5}}
\put(100,29.3){\line(-1,0){10}}
\put(120,36.3){\line(-1,0){21}}
\put(129,29.3){\line(-1,0){9}}
\put(90,22.3){\line(-1,0){5}}
\put(85,15.3){\line(1,0){10}}

\end{picture}

\end{center}

\vspace{5pt}

\noindent
The horizontal coordinate is $x^{1}$ and the vertical coordinate is $x^{2}$. The FZ particles are the bullets
joined by horizontal electric strings. The vertical electric flux consists of the FZ particles themselves. 
The lightest glueball is a pair of FZ particles with the same value of $x^{2}$. For 
small enough $g_{0}^{\prime}$, its mass is $2m_{1}$.

The leading-order vertical $k$-string tension is just the energy of the bound state of $k$ fundamental
FZ particles, divided by the lattice spacing \cite{k-string}. This yields a sine law
$\sigma_{\rm V}^{k}=\frac{m_{k}}{a} \propto \sin\pi k/N$. The leading-order horizontal $k$-string tension is
found by a simple argument \cite{k-string}. If 
the coupling $g_{0}^{\prime}$ is sufficiently small, then the 
mass gap in the principal chiral sigma models at $x^{2}=u^{2}$ and $x^{2}=u^{2}-a$ forces
electric flux along the line from $(u^{1},u^{2})$ to $(v^{1},u^{2})$. Thus the
potential is just that of the $(1+1)$-dimensional SU($N$) Yang-Mills theory, and
the horizontal string tension should behave as 
$\sigma_{\rm H}^{k}=\left( \frac{g_{0}^{\prime}}{a} \right)^{2}C_{k}$,
where $C_{k}$ is the quadratic Casimir operator. Adjoint sources are not confined \cite{k-string}.

These naive results for the string tension have further corrections in $g_{0}^{\prime}$, which were 
determined for the horizontal string tension for SU($2$) \cite{PhysRevD74}:
\beq
\sigma_{\rm H} =\frac{3}{2}\left( \frac{g_{0}^{\prime}}{a}\right)^{2} \left[ 1+
\frac{4}{3}\frac{0.7296}{C^{2}\pi^{2}}\frac{(g_{0}^{\prime})^{2}}{g_{0}^{2}}e^{4\pi/g_{0}^{2}} \right]^{-1}\;.
\label{string-tension}
\eeq
This calculation was done using the exact form factor for sigma model currents obtained by
Karowski and Weisz \cite{KarowskiWeisz}. Particle states have rapidity $\theta$ and four
internal states, labeled by $j$. The 2-particle current form factor is
\beq
\left< 0 \right\vert \!\!\!\!&\!\!\!j\!\!\!&\!\!\!^{L,R}_{0}(x)_{b} \left\vert \theta_{2},j_{2},\theta_{1},j_{1}\right>=
{\rm i}{\sqrt 2}\left( \delta_{j_{1} 4}\delta_{j_{2}b}  -
\delta_{j_{2} 4}\delta_{j_{1}b}  \pm \epsilon_{b j_{1} j_{2} }\right) 
m (\cosh \theta_{1} -\cosh \theta_{2}) \nonumber \\
\!\!&\!\!\times\!\!&\!\! 
\exp\{-{\rm i} m [ x^{0}(\cosh\theta_{1}+\cosh \theta_{2})-x^{1}(\sinh\theta_{1}+\sinh \theta_{2})
] \}\, F(\theta_{2}-\theta_{1}) \;, \label{useful-form-factor}
\eeq
where the plus or minus sign corresponds to the left-handed ($L$) or right-handed ($R$) current, respectively,
and
\beq
F(\theta)\!=\exp 2\int_{0}^{\infty} \frac{d\xi}{\xi}\, \frac{e^{-\xi}-1}{e^{\xi}+1}\,
\frac{\sin^{2} \frac{\xi(\pi{\rm i}-\theta)}{2\pi}}{\sinh \xi} 
\!=  \exp -\int_{0}^{\infty} \frac{d\xi}{\xi}\, \frac{e^{-\xi}}{\cosh^{2}\frac{\xi}{2}}\,
\sin^{2} \frac{\xi(\pi{\rm i}-\theta)}{2\pi}. \label{F-function}
\eeq
Other two-particle form factors can be obtained by crossing.

Our results for the mass gap and the string tension are not of the form one would expect for the isotropic
theory. If $g_{0}^{\prime}=g_{0}=e{\sqrt{a}}$, naive dimensional arguments imply that the gap is
proportional to $e^{2}$ and the string tension is proportional to $e^{4}$. In fact, general arguments
imply a crossover should occur as $g_{0}^{\prime}$ is increased, so this behavior is a real possibility 
as isotropy is approached \cite{PhysRevD74}. If this is so, there is no possibility of extracting the
isotropic gap and string tension from our results. Physical quantities cannot be
expected to have a part which is
analytic in both $e$ and $e^{\prime}$ (to see this, try to do standard perturbation theory in both
$e$ and $e^{\prime}$).

We have shown there is confinement
in the region $g_{0}$ and $g_{0}^{\prime}$ small, provided $g_{0}^{\prime}\ll g_{0}$. This leaves no doubt that
that confinement persists over the entire phase diagram of $g_{0}$ and $g_{0}^{\prime}$, even for the
isotropic case. Though 
we have little to say about the specifics of the isotropic theory, the anisotropic theory is perhaps
more interesting, as it is not simply finite, but asymptotically free. 

There are more problems to be investigated for anisotropic gauge theories. Corrections need
to be found in $g_{0}^{\prime}$ for the vertical string tension and the mass gap. The former
problem can probably only be solved easily for SU($2$). The theory should be studied at non-zero
temperature; it should be possible to see a phase transition to a deconfined phase. Matter fields
can be introduced and the spectrum of mesons and baryons should be examined. Finally, there
is a striking mathematical question. Our weak-coupling analysis looks very much like the strong-coupling
picture of a lattice gauge theory; is there a duality present?

\end{document}